\documentclass[doublecol,linenumbers]{epl2}

\title{Radio VLBI and the quantum interference paradox}
\shorttitle{Radio VLBI and the quantum interference paradox} 

\author{Ashok K. Singal}
\shortauthor{A. K. Singal}
\institute{Astronomy and Astrophysics Division, Physical Research Laboratory,
Navrangpura, Ahmedabad - 380 009, India.                    
Email: ashokkumar.singal@gmail.com 
}
\abstract{
We address here the question of interference of radio signals from astronomical sources like distant quasars, 
in a very long baseline interferometer (VLBI), where two (or more) distantly located radio telescopes (apertures),  
receive simultaneous signal from the sky. In an equivalent optical two-slit experiment, it is 
generally argued that for the photons involved in the interference pattern on the screen, it is not possible, even in principle, to 
know which of the two slits a particular photon went through and that any attempt to ascertain this 
destroys the interference pattern. But in the case of the modern radio VLBI, it is a routine matter to record the 
phase and amplitude of the voltage outputs from the two radio antennas on a recording media separately and then do 
the correlation between the two recorded signals later in an offline manner. 
Does this not violate the quantum interference principle? We provide a resolution of this problem here.
}
\begin{document}
\maketitle
\section{Introduction}
According to the quantum interference principle, in Young's two-slit optical interference 
experiment, a photon cannot be treated as a classical discrete particle and for all purposes it seems to be 
passing through both the slits \cite{22.1,22.2,22.3}.
It is supposed that for any photon that appears on the screen,  
giving rise to the interference pattern, it is not possible (even in 
principle) to state which particular slit the photon went through. Any attempt
to detect the path of the photon trajectory (say, by putting appropriate sensors 
at one or both slits) will destroy the interference pattern. Now radio 
Very Long Baseline Interferometry (VLBI), where two (or more) distantly located radio telescopes (apertures),
which could even be on different continents or while one of them is on Earth the other antenna could be in space, 
receive simultaneous signal from the sky, apparently seems to routinely provide a mechanism whereby one 
records the phases and amplitudes of the radio waves arriving at 
each of the antennas independently. Since the technique does not destroy the phases 
of the incoming waves while recording their signature, one regularly
gets the equivalent of Young's  interference pattern by correlating
the signal from both antennas in an offline manner by introducing a variable additional phase shift between
the two voltage signals. It looks like one is able to overcome the quantum limitation
in this case as one is able to record the presence of the radio photon at each slit
(by getting its phase and amplitude at individual antennas) without
destroying the interference pattern. Is there a breakdown of the
quantum interference principle here?
\section{An issue of sensitivity}
This interesting question, with reference to radio VLBI was first
raised by Burke \cite{22.4,22.5}. Burke's solution to this paradox was that
the detection of an individual photon at either of the antenna
would require a very sensitive system consisting of one or more
radio signal amplifiers and that it may not really be possible to
build such sensitive receivers that could detect the presence of
individual photons. In his opinion, the paradox is resolved by
postulating the absence of such sensitive detectors at each
individual antenna (even in principle). We may add here that 
radio astronomers do employ auto correlator systems where 
they successfully detect signals from single antennas too. Of course 
signals from individual ``single'' radio photons is never an option 
in radio astronomy, as individual single are detectable neither in an  
autocorrelation (from a single telescope) nor in a cross-correlator mode 
(with signals from two or more telescopes).
\section{Classical or quantum picture}
Radio astronomers almost exclusively use only the classical wave 
picture of electromagnetic radiation to describe or interpret their 
observations. They hardly ever resort to the quantum mechanical picture 
of radio photons to describe the phenomena observed. But both ways 
of description should be consistent  \cite{22.3,22.6}. For example the 
total intensity interferometer \cite{22.7} describes the picture at 
optical frequencies in quantum mechanical way. But the same phenomenon 
manifests itself in classical wave picture where the intensity fluctuations in the noise 
(after it has gone through an auto-correlation detector) is found to cross-correlate 
in different simultaneous beams of a correlator system in the same way as 
a signal from a source in sky in different beams  \cite{22.8}.
\section{Optical versus radio techniques}
We want to show here that the solution of the paradox does not lie
in the technical question of whether or not such sensitive amplifiers, 
which can help detect individual photons, are technically feasible or
not. It is more of a question of principle. In fact, to us the argument
about the amplifiers appears rather extraneous to the central issue
here. After all individual radio photons are not detected even in 
the cross-correlator output. Our main aim in the two slit experiment is not 
{\em to detect} individual photons at each slit that later may be taking part 
in interference pattern; our interest in the present context instead is only 
to know if that photon has passed through that slit, e.g., using a recoil system 
which may detect the passage of the photon without annihilating it. 
In radio case we can detect the photon without annihilating it, as we can make copies 
of the recorded signal without destroying or tempering with it. The actual physical 
detection of a photon at an aperture or the mere knowledge of the passage of the photons 
through that aperture are two quite distinct phenomena.

There is a fundamental difference in the way the optical and radio 
techniques are applied, even though the basic physics involved is the same. 
In Young's two-slit experiment there is a screen or set of detectors (after the slits)
where an independent detection of the arrived photon is made (on a
photographic plate or on some other detector). 
The question of the path taken by the photon arises only after a photon has already 
been detected. The conventional wisdom is that a
photon detected on the screen in the optical two slit case, for all purposes passes 
through both slits and that we cannot trace its path through a single slit alone 
without destroying the interference pattern. 
However, in the case of radio VLBI, the detection of the photon itself is 
through the correlation output of the {\em both} slits. In other words, unless we positively 
find a correlated signal from both the antennas (irrespective of whether we do an on-line or 
offline correlation between the two), we do not even know that a photon is present 
in our system and a question that which individual slit (or antenna
in this case) it passed through does not even arise.
In optical case where there is an independent detection of the photon, 
one might legitimately pose the question - which slit it has gone through? 
But in radio VLBI, there is no ambiguity about this and 
a detection at each slit is not even necessary (irrespective of whether or not 
it is even possible because of high noise-figure of amplifiers/detectors) 
as we have knowledge about that and our anxiety about which slit the photon has passed through 
is already a gone conclusion. Therefore the further question of detectability of the photon 
at either of the `radio slits' (antennas) is not even meaningful.

Even otherwise, suppose one were to detect individual photons at either of the antennas 
(somehow bypassing Burke's limit on possible detector systems), these photons could be
coming from anywhere (non-coherent arrival of photons at the two antenna
from any other sources of radiation). Such photons are not
of any interest to us as such in any case are plenty adding to our
system noise. There is no guarantee that these are the same set of photons 
that give rise to the correlation in the cross-correlated output, as 
there is plenty of noise accompanying or superimposed on the correlated signal 
and it is not possible to know which particular photons, even if detected on 
individual VLBI sub-station records, are the ones that give rise to correlation 
between the two stations. In radio VLBI case there is no independent information
about the arrival of a photon taking part in the interference pattern
except by a positive correlation of signal at both antennas. This is
quite in contrast to the optical case where the presence of such a
photon is detected independent of its presence noted at individual 
slits. 
\section{Single radio-photon interference}
While in the case of an optical double-slit experiment, the photons coming 
are only our considered source of coherent emission, in the case of radio VLBI experiments, 
there is in general a large flux of radio photons from all known/unknown sources in 
different directions, like the sky background, ground reflection, instrument noise and many 
other such `unwanted' causes. In fact the only way to find or ascertain that a radio photon 
is indeed coming from the source of our interest (say, compact radio core of a quasar) 
is from a correlated output from both VLBI sub-stations. Any correlated signal in this 
offline done analysis is immediately ascribed to our quasar, and if a correlated signal is not seen, 
we say that nothing has been detected from the quasar. Thus even if Burke's difficulties of detecting 
photons from individual antennas are overcome and one does detect some positive output from one or both 
VLBI sub-stations, it will remain unknown whether the detected signal is from the quasar or from elsewhere, 
as presence of a photon from the quasar is registered only through a correlated output seen in offline analysis. 
Each sub-station receives copious amount of radio photons from sources all over, even when we are not 
receiving any signal from the coherent source of our interest, like the distant quasar. 

Like in the optical two-slit experiments, where the interference pattern is seen even when the incident light intensity is reduced so much that at a 
given instant only a single photon could be deemed to be present in the experiment \cite{22.1,22.2,22.13,22.14}, one could try to ascertain that if such a thing 
could happen even in the radio VLBI experiments where usually one assumes a huge number (millions!) of radio photons to be 
simultaneously present in the signal. 
For example a typical source signal in radio VLBI observations is about one Jansky (Jy) \cite {22.12}, 
where a Jy stands for $10^{-26}$ Watts m$^{-2}$ Hz$^{-1}$. 
The total amount of radio power collected by a radio antenna of effective aperture (area) $A$ and observing at a frequency $\nu$ with a bandwidth 
$\Delta\nu$ is then, $P= 10^{-26} A \Delta\nu$. Now the energy carried by each radio photon is $h \nu$, where $h=6.626 \times 10^{-34}$ Joules-sec is the 
Planck's constant. Then the temporal rate of number of photons collected by the antenna is $10^{-26} A \Delta \nu/h \nu \approx 1.5 \times 10^{7} A \Delta \nu/\nu$ 
s$^{-1}$. Now coherence length for a signal of bandwidth $\Delta\nu$ is $c/\Delta\nu$, which effectively is the longitudinal 
extent of the photon \cite{22.2,22.6}, or equivalently the temporal duration of the photon is $1/\Delta\nu$ (from 
Heisenberg's uncertainty principle). Then during this interval of time the number of photons falling on the antenna is 
${\cal N} \sim 1.5 \times 10^{7} (A \Delta \nu/\nu) \times (1/\Delta \nu)=1.5 \times 10^{7} A/\nu$. For a typical VLBI experiment 
employing paraboloids \cite {22.12} of diameter $D=25$m, with physical area $A=\pi (25/2)^2 \approx 500$ m$^2$, and $\nu= 43$ GHz, 
we get ${\cal N} \sim 0.17$, which is of the order of unity. Therefore we can say that at least in such cases the number of photons 
falling at any instant (within the coherence length time) is of the order of unity even in radio VLBI experiments.

We should reiterate that even if number of photons falling at any instant may be of the order of unity, radio telescopes do not 
ever detect single photons either in the individual antennas (auto correlator systems) or in interferometer pairs 
(cross-correlator systems). In general, the radio signal is integrated over a time interval $\tau\sim$ tens of seconds, 
it amounts to integrating over ${N} \sim \Delta \nu\: \tau$  photons and thereby improving the signal to noise by a 
factor $\sim \sqrt {N}$. With a typical $\Delta \nu \sim$ 10 MHz, the improvement in signal to noise can be 
$\sqrt {\Delta \nu\: \tau}\sim 10^4$.

Even if we take the view that a photon cannot be deemed to exist at a pre-detection stage,
in the case of radio VLBI, the recording on tapes at individual sub-stations is as good as a 
post-detection stage, as no further changes in the detection probability are possible. In optical two-slit experiment 
the photon picture comes into existence only at detection time when the probability wave function 
``collapses'' or gets frozen in the sense that there are no further temporal changes in it. But this 
is what happens in the radio VLBI case at each sub-station as we have `permanent' records on media, 
which are not going to change and thus the data recorded is `frozen' in time, with probability wave 
function not going to change or `collapse' any further.  

Now a question could be raised. Is it `one and the same' photon whose presence has got registered on tapes at both VLBI stations which are 
far apart say, 8 to 10 thousand kilometres (across the diameter of earth)? At least this seems to be in accordance with the conventional wisdom. 
In fact with the space VLBI experiments being a success \cite{22.10,22.11}, the distances could be even larger. 
As different photons are not having any coherent phase relation with each other (except perhaps in a specially prepared system like in a LASER), 
that means a positive correlation implies the presence of one and the same photon at two antennas thousands of km apart. 
\section{Conclusions}
We showed that there is no breakdown of the quantum interference principle in the case of radio VLBI, where the 
phase and amplitude of the signal from a common source in sky, falling on two (or more) distantly located 
radio telescopes (apertures), are recorded separately, and the correlation between the two recorded signals 
is done later in an offline manner. The argument put forward in an equivalent optical two-slit experiment is 
that it is not possible, even in principle, 
to ascertain which of the two individual slits a particular photon involved in the interference pattern has gone through and that any procedure to ascertain this would 
destroy the interference pattern. In the past literature a solution to this paradox in the case of radio VLBI has been offered by saying 
that it may not really be possible to build such sensitive radio receivers that could detect the presence of
photons from individual antennas. We argued that in radio VLBI case, the detection of the photon itself is 
through the correlation output of the {\em both} slits (antennas)
and there is no violation of the quantum interference principle.

\end{document}